\documentstyle[editedvolume,NATO]{crckapb} 

\begin{opening}
\title{PKS2250-41 and the role of jet cloud interactions in powerful radio galaxies.}
\subtitle{}

\author{M.Villar-Mart\'\i n}
\institute{Dept. of Physics, University of
Sheffield\\ Sheffield S3~7RH, UK}

\author{C.Tadhunter}
\institute{Dept. of Physics, University of
Sheffield\\ Sheffield S3~7RH, UK}

\author{R. Morganti}
\institute{Istituto di Radioastronomia\\ Via Gobetti 101, 40129 Bologna, 
Italy}

\author{D. Axon}
\institute{Dept. of Physics and Astronomy, University of Manchester\\ Oxford
Road, Manchester, UK}

\author{A. Koekemoer}
\institute{Space Telescope Science Institute\\ 3700 San Martin
Drive, Baltimore, MD21218, USA}
 
\end{opening}

\runningtitle{Jet cloud interactions in powerful radio galaxies.}

\begin{document}


\section{Abstract}
	
	We have studied the effects of the interaction between the radio
jet and the ambient gas in the powerful radio galaxy PKS2250-41 ($z=$0.31). Our
results show that the   gas has been accelerated, compressed, heated and collisionally ionized by the shock. This study helps us to understand the processes which
determine the observed properties of many high redshift radio galaxies.
 
\section{Introduction}
 
	The observed properties of radio galaxies  at $z>$0.7 are very 
different to those at low redshift. The extended optical structures (line emission and continuum) are closely
aligned with the radio axis and often highly collimated. 
 Although several explanations
are popular to explain this alignment effect [1,2], the interaction between the radio
jet and the ambient gas is certainly playing a role.   In spite of the much
observational  evidence for such interactions, not only in distant
radio galaxies, but
also in some closer radio galaxies and Seyferts, the physics involved is
not well understood. We are carrying out  a program to study a sample of low-intermediate
redshift ($z<$0.4) radio galaxies undergoing jet-cloud interactions, in
order to study
in detail how the interaction happens and its consequences for  the observed
properties of the objects.

	We present here results of the study of   PKS2250-41,
a radio galaxy in which a very strong interaction
is taking place between the radio jet and an arc-like structure to the W of
the main galaxy [3,4]. 

\section{Observations}

	Optical spectra were obtained  with the RGO
spectrograph at AAT (AAO). The slit was located at different positions parallel
and perpendicular to the radio axis so that the whole arc (the region
where the interaction is strongest) was covered. By fitting the emission lines,
we have isolated the different kinematic components
 and have studied their kinematic, flux and physical properties. Our goal is to discriminate   
the dominant physical mechanism for each gaseous component.

\section{Results}

\centerline{\bf Kinematics and ionization level}

	The kinematics are complex across the arc. The [OIII]5007,4959
and H$\beta$ lines split into  several kinematic components for which FWHM, flux
and [OIII]5007/H$\beta$ were studied. The two main
components are:

A {\it broad component} characterized by:

\begin{itemize}

\item morphological association with the   radio lobe;

\item large line widths (FWHM$\sim$500-900 km s$^{-1}$);

\item line width and flux peak at the position of the radio hot spot;

\item low ionization level ([OIII]/H$\beta\sim$1-3). 

\end{itemize}

A {\it narrow component} characterized by: 

\begin{itemize}

\item its presence across the whole surface of the arc and far beyond the edge
of the radio lobe;

\item  small line widths (FWHM$\sim$60-200  km s$^{-1}$);

\item  relatively high ionization level ([OIII]/H$\beta\sim$4-9);

\end{itemize}

\centerline{\bf Electronic Temperature}

	We have used the measured    $\frac{[OIII]\lambda5007}{[OIII]\lambda4363}$ ratios
to calculate the electronic temperatures for the narrow and broad components
[5].
 The equivalent width of
the [OIII]$\lambda$4363 line is large in the regions of interest and
systematic effects related to the subtraction of the
underlying stellar continuum are not a significant source
of error. 

	We measure $\frac{[OIII]\lambda5007}{[OIII]\lambda4363}$=
50$\pm$20  for the narrow component and 13$\pm$4 for the broad component, which
yield T$_e$= 15,000$^{+1300}_{-1000}$ and    41,000$^{+10,000}_{-8,000}$ 
respectively.  Therefore, {\it the gas emitting the broad component is hotter}.

\centerline{\bf Density}

	 We have used  the [SII]$\lambda\lambda$6716,6731  
doublet to calculate the electronic densities for the narrow and broad components [5].  
We have assumed the T$_e$ values calculated in the previous section. 	We measure $\frac{[SII]\lambda6716}{[SII]\lambda6731}$ = 1.47$\pm$0.19 for the narrow component and  1.13$\pm$0.08 for the broad
component, which yield  n$_e<$120 and   n$_e$=570$^{+260}_{-190}$
respectively. Therefore,  {\it the gas emitting the broad component is denser}.

\centerline{\bf Ionization mechanism}Two main mechanisms   can  ionize the gas:
1) AGN (active nucleus) photoionization 2) Shocks via    collisional ionization
of the gas and further radiative cooling and via the generation of a strong 
ionizing UV continuum.  Which mechanism  is responsible for the emission of
the narrow and the broad components resolved in PKS2250-41?
 
	The diagnostic diagram [OIII]$\lambda$5007/$\lambda$4363 vs. HeII/H$\beta$ provides a strong method to discriminate between 
AGN  
and shock models [3].  We have compared  the measured values for the line ratios of
the broad and narrow components with the model predictions (see Fig.~1).    Notice the clear separation between the two kinematic components due to
the low [OIII]$\lambda$5007/$\lambda$4363 (high T$_e$) and faintness of the
HeII emission of the broad component. The AGN photoionization models  
 predict too high
[OIII]5007/4363 ratios ({\it i.e.}  low electron
temperatures) and do not appear in the diagram. Adding matter bounded models  
can make the position of the narrow component  consistent with AGN models [6]. However these cannot explain the line ratios of the broad component, which on the other
hand is consistent with shock (radiative cooling) models. Therefore, the line ratios
of the broad component suggest that    
it  {\it represents collisionally ionized gas which is
cooling behind a shock front}. The line ratios of the {\it narrow component 
suggest that it has been ionized by a strong UV continuum} (AGN and/or shock
generated).

\begin{figure}  
\includegraphics{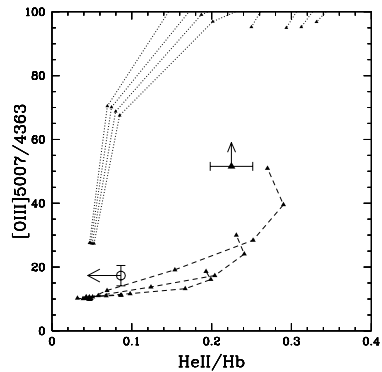}
\vspace{2in}
\caption{[OIII]$\lambda$5007/[OIII]$\lambda$4363
{\it vs.}  HeII/H$\beta$.  The broad   (open circle) and the narrow
(solid triangle) components are clearly separated in this diagram. 
 Dashed lines: shock models (shocked
gas cooling
radiatively). Dotted lines: Shock models (adding  the contribution of
ionization by the UV shock continuum).}
\end{figure} 

\section{Discussion}
	
		Hydrodynamical models for an advancing jet in a gaseous medium show
that the shocked gas is compressed and accelerated and the temperature
raised. The gas will be collisionally
ionized  and will then cool   radiatively.  

	All the properties of the broad component are
consistent with it being shocked gas: Higher temperature and density
--$>$ heating and compression.
Low ionization level --$>$ compression and/or shock ionization. High
velocities --$>$acceleration. Faint HeII and high temperature --$>$ shock ionization. On the other hand,
the   narrow component  is probably ambient gas whose properties can be explained in terms of
pure gravitational motions and photoionization by a 
hard continuum source. This gas has not interacted with the radio jet.

\section{Conclusions}

We have studied the effects of the interactions between the radio jet and
the ambient gas in PKS2250-41. The results provide clues to understanding
the observed properties of many distant
radio galaxies.
   
We have resolved  kinematically the emission from the gas interacting
with the radio jet and the emission from the non shocked ambient gas.
The very different properties of these two gaseous components  
show the important effects of the interaction on the physical, morphological,
ionization and kinematic properties of the gas.

\end{document}